# Nonlinear wave solution to a coupled mKdV equations with variable coefficients

Wenjuan Wu

College of Mathematics and Statistics, Chongqing University, Chongqing, PR China

**Abstract:** The nonlinear wave solutions to coupled mKdV equations with variable coefficients are obtained by using the F-expansion method, including 12 kinds of Jacobi elliptic function solutions. In the limit cases, the torsional wave solutions, periodic solutions and solitary wave solutions are obtained as well.

## 1. Introduction

As an important mathematical physics model, KdV equation [1] has been widely concerned and studied by scholars since it was proposed. It was first generalized to modified KdV equations [2], and further developed into higher-order KdV equations [3] and even combined or coupled equations [4]. Moreover, compared with the KdV equation with constant coefficient, the KdV equation with variable coefficient has more practical application value [5]. In recent years, the coupled KdV equations with variable coefficients have received extensive attention. In recent years, the variable coefficient coupled KdV equation has received extensive attention, and the exact solution of them has obtained a lot of results[6].

From the perspective of extending KdV equation to mKdV equation and adding coupling terms, we propose the coupled mKdV equations with variable coefficients in the following form

$$\begin{cases} u_t + a(t)u^2 u_x + b(x)v^2 v_x + c(t)u_{xxx} = 0 \\ v_t + a(t)u^2 v_x + d(t)v^2 u_x + c(t)v_{xxx} = 0 \end{cases} \quad (1.1)$$

where $a(t), b(t), c(t), d(t)$ are all nonzero functions of variable $t$ only, and satisfy the following conditions

$$\frac{e(t)}{b(t)} = \sigma^2, \quad \frac{c(t)}{a(t) + \sigma^3 b(t)} = k, \quad \sigma, k \ are\ cons\tan ts. \quad (1.2)$$

In this Letter, We use F-expansion method to solve Equations (1.1) with conditions(1.2).

## 2 Description of F-expansion method

For nonlinear wave equations

$$P(u, u_t, u_x, u_{tt}, u_{xt}, u_{xx}, u_{xxx}, \ldots) = 0 \tag{2.1}$$

Where P is a polynomial function about $u, u_t, u_x, u_{tt}, u_{xt}, u_{xx}, u_{xxx}$ and so on, solving the equation by F-function expansion method can be divided into the following 5 steps.

Step1. Find the traveling wave solution of Equation (2.1) in the following form

$$u(x,t) = f(\xi), \xi = \lambda x + \mu(t) \tag{2.2}$$

where $\lambda$ is a nonzero constant, $u(t)$ is a nonzero function of variable $t$ only satisfying $\mu'(t) = \omega\bar{\mu}(t)$. Substitute (2.2) into (2.1) to get the ordinary differential equation about $f(\xi)$

$$\tilde{P}(f, \mu'f', \lambda f', \mu''f', \mu'^2 f'', \mu'\lambda f'', \lambda^2 f'', \lambda^3 f''', \ldots) = 0 \tag{2.3}$$

or

$$\tilde{P}(f, \omega\bar{\mu}f', \lambda f', \omega\bar{\mu}'f', (\omega\bar{\mu})^2 f'', \omega\bar{\mu}\lambda f'', \lambda^2 f'', \lambda^3 f''', \ldots) = 0 \tag{2.4}$$

Step2. Suppose $f(\xi)$ can be expanded as a finite power series of $F(\xi)$

$$f(\xi) = \sum_{k=0}^{n} a_k F^k(\xi), a_n \neq 0 \tag{2.5}$$

where $a_0, a_1, \ldots, a_n$ are constants to be determined later, and the positive integer $n$ can be obtained by balancing the highest degree of the nonlinear term with the highest degree of the highest derivative term in (2.3). $F(\xi)$ satisfy the first order nonlinear ordinary differential equation

$$F'^2(\xi) = q_0 + q_2 F^2(\xi) + q_4 F^4(\xi) \tag{2.6}$$

and further we have

$$\begin{cases} F'F'' = q_2 FF' + 2q_4 F^3 F' \\ F'' = q_2 F + 2q_4 F^3 \\ F''' = q_2 F' + 6q_4 F^2 F' \end{cases} \tag{2.7}$$

So the relationship between $q_0, q_2, q_4$ and $F(\xi)$ is as follows:

Table 2.1 The relationship of $q_0, q_2, q_4$ and $F$

| $q_0$ | $q_2$ | $q_4$ | $F'^2 = q_0 + q_2 F^2 + q_4 F^4$ | $F$ |
|---|---|---|---|---|
| 1 | $-(1+m^2)$ | $m^2$ | $F'^2 = (1-F^2)(1-m^2 F^2)$ | $sn\xi, cd\xi = \dfrac{cn\xi}{dn\xi}$ |
| $1-m^2$ | $2m^2-1$ | $-m^2$ | $F'^2 = (1-F^2)(m^2 F^2 + 1 - m^2)$ | $cn\xi$ |
| $m^2-1$ | $2-m^2$ | $-1$ | $F'^2 = (1-F^2)(F^2 + m^2 - 1)$ | $dn\xi$ |
| $m^2$ | $-(1+m^2)$ | 1 | $F'^2 = (1-F^2)(m^2 - F^2)$ | $ns\xi = (sn\xi)^{-1}, dc\xi = \dfrac{dn\xi}{cn\xi}$ |
| $-m^2$ | $2m^2-1$ | $1-m^2$ | $F'^2 = (1-F^2)[(m^2-1)F^2 - m^2]$ | $nc\xi = (cn\xi)^{-1}$ |
| $-1$ | $2-m^2$ | $m^2-1$ | $F'^2 = (1-F^2)[(1-m^2)F^2 - 1]$ | $nd\xi = (dn\xi)^{-1}$ |
| 1 | $2-m^2$ | $1-m^2$ | $F'^2 = (1+F^2)[(1-m^2)F^2 + 1]$ | $sc\xi = \dfrac{sn\xi}{cn\xi}$ |
| 1 | $2m^2-1$ | $-m^2(1-m^2)$ | $F'^2 = (1+m^2 F^2)[(m^2-1)F^2 + 1]$ | $sd\xi = \dfrac{sn\xi}{dn\xi}$ |
| $1-m^2$ | $2-m^2$ | 1 | $F'^2 = (1+F^2)(F^2 + 1 - m^2)$ | $cs\xi = \dfrac{cn\xi}{sn\xi}$ |
| $-m^2(1-m^2)$ | $2m^2-1$ | 1 | $F'^2 = (F^2 + m^2)(F^2 + m^2 - 1)$ | $ds\xi = \dfrac{dn\xi}{sn\xi}$ |

Step3. Substitute the expansion (2.5) into the ordinary differential equation (2.3) of $f(\xi)$. Then using formula (2.6), (2.7), the left side of formula (2.3) can be converted to a polynomial of $F$. Next let the coefficients of the polynomial be 0, the algebraic equations of $a_0, a_1, \ldots, a_n, \omega, \lambda$ can be obtained.

Step4. Use Mathematica software or Wu elimination method to solve the algebraic equations obtained in Step3. Representing $a_0, a_1, \ldots, a_n, \omega, \lambda$ by $(q_0, q_2, q_4)$, and substituting the result into the expansion (2.5), a general form of a traveling wave solution of the equations (2.1) can be obtained.

Step5. Select the appropriate $(q_0, q_2, q_4)$ in (2.6) to make $F$ the Jacobi elliptic function. The correspondence between the value of $(q_0, q_2, q_4)$ and the Jacobi elliptic functions has been given in Table 2.1, and there are 12 kinds of correspondence. Select $(q_0, q_2, q_4)$ and Jacobi elliptic function solution $F$ in the table and substitute them into the general form solution of (2.1) in Step4, then the Jacobi solution elliptic function of (2.1) will be obtained.

## 3 Nonlinear wave solutions

Now we solve equations (1.1) via F-expansion method.

Step1. Take

$$u = f(\xi), \quad v = g(\xi), \quad \xi = \lambda x + \mu(t), \quad \lambda \neq 0 \qquad (3.1)$$

where $\lambda$ is a nonzero constant and $\mu(t)$ determined later is a nonzero function of $t$. Substituting equations (3.1) into equations (1.1) yields an ordinary differential equations

$$\begin{cases} \mu'f' + \lambda a f^2 f' + \lambda b g^2 g' + \lambda^3 c f''' = 0 \\ \mu'g' + \lambda a f^2 g' + \lambda d g^2 f' + \lambda^3 c g''' = 0 \end{cases} \quad (3.2)$$

Suppose equations (3.2) has a solution of the following form:

$$\begin{cases} f(\xi) = \sum_{i=0}^{n} a_i F^i(\xi), \; a_n \neq 0 \\ g(\xi) = \sum_{i=0}^{m} b_i F^i(\xi), \; b_m \neq 0 \end{cases} \quad (3.3)$$

where $a_1, a_2, \ldots, a_n$ and $b_1, b_2, \ldots, b_m$ arecinstans to be determined later and $F(\xi)$ satisfys (2.6).

Next determine the values of n and m by balancing the highest degree of $f^2 f'$, $g^2 g'$ and $f'''$ (or $f^2 g'$, $g^2 f'$ and $g'''$), we have

$$O(f^2 f') = 3n + 1, \; O(g^2 g') = 3m + 1, \; O(f''') = n + 3.$$

Making $O(f^2 f') = O(g^2 g') = O(f''')$, we obtain

$$m = n = 1. \quad (3.4)$$

Substituting equation (3.4) into equations (3.3) leads to the following equations:

$$\begin{cases} f(\xi) = a_0 + a_1 F(\xi) \\ g(\xi) = b_0 + b_1 F(\xi) \end{cases} \quad (3.5)$$

Now substituting (3.5) into (3.2) yields

$$\begin{cases} a_1 \mu' F' + \lambda a (a_0^2 a_1 F' + 2 a_0 a_1^2 F F' + a_1^3 F^2 F') \\ + \lambda b (b_0^2 b_1 F' + 2 b_0 b_1^2 F F' + b_1^3 F^2 F') + \lambda^3 c (a_1 q_2 F' + 6 a_1 q_4 F^2 F') = 0 \\ b_1 \mu' F' + \lambda a (a_0^2 b_1 F' + 2 a_0 a_1 b_1 F^2 F' + a_1^2 b_1 F^2 F') \\ + \lambda d (a_1 b_0^2 F' + 2 a_1 b_0 b_1 F F' + a_1 b_1^2 F^2 F') + \lambda^3 c (b_1 q_2 F' + 6 b_1 q_4 F^2 F') = 0 \end{cases}$$

Canceling $F'$, we get further

$$\begin{cases} (a_1 \mu' + \lambda a_0^2 a_1 a + \lambda b_0^2 b_1 b + \lambda^3 a_1 q_2 c) + 2\lambda (a_0 a_1^2 a + b_0 b_1^2 b) F \\ + \lambda (a_1^3 a + b_1^3 b + 6\lambda^2 a_1 q_4 c) F^2 = 0 \\ (b_1 \mu' + \lambda a_0^2 b_1 a + \lambda a_1 b_0^2 d + \lambda^3 b_1 q_2 c) + 2\lambda (a_0 a_1 b_1 a + a_1 b_0 b_1 d) F \\ + \lambda (a_1^2 b_1 a + a_1 b_1^2 d + 6\lambda^2 b_1 q_4 c) F^2 = 0 \end{cases} \quad (3.6)$$

Setting each coefficient of $F^k$ ($k = 0, 1, 2$) to zero yields an algebraic equations of

$a_0, a_1, b_0, b_1, \lambda$ and $\omega$

$$a_1 \mu' + \lambda a_0^2 a_1 a + \lambda b_0^2 b_1 b + \lambda^3 a_1 q_2 c = 0 \tag{3.7}$$

$$a_0 a_1^2 a + b_0 b_1^2 b = 0 \tag{3.8}$$

$$a_1^3 a + b_1^3 b + 6\lambda^2 a_1 q_4 c = 0 \tag{3.9}$$

$$b_1 \mu' + \lambda a_0^2 b_1 a + \lambda a_1 b_0^2 d + \lambda^3 b_1 q_2 c = 0 \tag{3.10}$$

$$a_0 a_1 b_1 a + a_1 b_0 b_1 d = 0 \tag{3.11}$$

$$a_1^2 b_1 a + a_1 b_1^2 d + 6\lambda^2 b_1 q_4 c = 0 \tag{3.12}$$

The solution of equations (3.7)-(3.12) under condition (1.2) and

$$a_1 \neq 0, \ b_1 \neq 0, \ \lambda \neq 0$$

is only

$$b_0 = -\frac{a(t)}{d(t)} a_0, \quad a_1 = \pm \sqrt{-6\lambda^2 k q_4}, \quad b_1 = \pm \sigma \sqrt{-6\lambda^2 k q_4}$$

$$\mu = -\int_0^t \lambda a_0^2 a(\tau) + \lambda \sigma a_0^2 \frac{a^2(\tau) b(\tau)}{d(\tau)} + \lambda^3 q_2 c(\tau) d\tau, \tag{3.13}$$

$$\lambda, \ a_0 = \text{arbitrary constant}.$$

Substituting (3.13) into (3.5) yields general form solutions to (1.1),(1.2)

$$\begin{cases} f(\xi) = a_0 \pm \sqrt{-6\lambda^2 k q_4}\, F(\xi) \\ g(\xi) = -\dfrac{a(t)}{d(t)} a_0 \pm \sigma \sqrt{-6\lambda^2 k q_4}\, F(\xi) \end{cases} \tag{3.14}$$

with $\xi = \lambda x - \int_0^t \lambda a_0^2 a(\tau) + \lambda \sigma a_0^2 \dfrac{a^2(\tau) b(\tau)}{d(\tau)} + \lambda^3 q_2 c(\tau) d\tau$.

Taking $q_0 = 1$, $q_2 = -(1 + m^2)$, $q_4 = m^2$, we have

$$\begin{cases} f(\xi) = a_0 \pm \sqrt{-6k\lambda^2 m^2}\, sn(\xi) \\ g(\xi) = -\dfrac{a(t)}{d(t)} a_0 \pm \sigma \sqrt{-6k\lambda^2 m^2}\, sn(\xi) \end{cases} \tag{3.15}$$

with $\xi = \lambda x - \int_0^t \lambda a_0^2 a(\tau) + \lambda \sigma a_0^2 \dfrac{a^2(\tau) b(\tau)}{d(\tau)} - (1 + m^2) \lambda^3 c(\tau) d\tau$.

In the limit case when $m \to 1$, (3.15) becomes

$$\begin{cases} f(\xi) = a_0 \pm \sqrt{-6\lambda^2 k}\,\tanh(\xi) \\ g(\xi) = -\dfrac{a(t)}{d(t)}a_0 \pm \sigma\sqrt{-6\lambda^2 k}\,\tanh(\xi) \end{cases} \quad (3.16)$$

with $\xi = \lambda x - \int_0^t \lambda a_0^2 a(\tau) + \lambda\sigma a_0^2 \dfrac{a^2(\tau)b(\tau)}{d(\tau)} - 2\lambda^3 c(\tau) d\tau,$

which is torsional wave solution of Equations(1.1),(1.2).

Taking $q_0 = 1-m^2$, $q_2 = 2m^2 -1$, $q_4 = -m^2$, we obtain the periodic wave solution expressed by Jacobi cn-function to Equations(1.1),(1.2)

$$\begin{cases} f(\xi) = a_0 \pm \sqrt{6k\lambda^2 m^2}\,cn(\xi) \\ g(\xi) = -\dfrac{a(t)}{d(t)}a_0 \pm \sigma\sqrt{6k\lambda^2 m^2}\,cn(\xi) \end{cases} \quad (3.17)$$

with $\xi = \lambda x - \int_0^t \lambda a_0^2 a(\tau) + \lambda\sigma a_0^2 \dfrac{a^2(\tau)b(\tau)}{d(\tau)} + (2m^2-1)\lambda^3 c(\tau) d\tau.$

In the limit case when $m \to 1$, (3.17) becomes

$$\begin{cases} f(\xi) = a_0 \pm \sqrt{6k\lambda^2}\,\mathrm{sech}(\xi) \\ g(\xi) = -\dfrac{a(t)}{d(t)}a_0 \pm \sigma\sqrt{6k\lambda^2}\,\mathrm{sech}(\xi) \end{cases} \quad (3.18)$$

with $\xi = \lambda x - \int_0^t \lambda a_0^2 a(\tau) + \lambda\sigma a_0^2 \dfrac{a^2(\tau)b(\tau)}{d(\tau)} + \lambda^3 c(\tau) d\tau,$

which is solitary wave solution of Equations(1.1),(1.2).

## 4 Conclusion

The problem of solving couple mKdV equations with variable coefficients is transformed into algebraic equations by using F-expansion method. By limiting the coefficients, the general solutions of the varied-coefficient coupled mKdV equations are obtained. In the limit case, we also obtain the torsional wave solutions, periodic wave solutions and isolated wave solution.